\documentclass[5p,times]{elsarticle}
\usepackage{ecrc_CPC}

\usepackage[T1]{fontenc}
\usepackage{hyperref}
\usepackage{pdfpages}
\usepackage{multirow}  
\usepackage{amsmath,amssymb,bbm,braket}
\usepackage{physics}
\usepackage{caption,subcaption}  
\usepackage{orcidlink}

\usepackage{caption}
\usepackage{subcaption}

\volume{xxx}
\firstpage{1}
\journalname{Computer Physics Communications}
\runauth{Au-Yeung et al.}
\jid{CPC}
\biboptions{comma,square}  

\begin{document}

\begin{frontmatter}

\title{Quantum Algorithm for Smoothed Particle Hydrodynamics}

\author[strath]{R. Au-Yeung\corref{cor1}}
\ead{rhonda.au-yeung@strath.ac.uk}

\author[manc]{A. J. Williams}

\author[strath]{V. M. Kendon\corref{cor1}}
\ead{viv.kendon@strath.ac.uk}

\author[manc]{S. J. Lind\corref{cor1}}
\ead{steven.lind@manchester.ac.uk}

\address[strath]{Department of Physics, University of Strathclyde, United Kingdom}
\address[manc]{School of Engineering, University of Manchester, United Kingdom}

\cortext[cor1]{Corresponding authors:}

\begin{abstract}
We present a quantum computing algorithm for the smoothed particle hydrodynamics (SPH) method. We use a normalization procedure to encode the SPH operators and domain discretization in a quantum register. We then perform the SPH summation via an inner product of quantum registers. Using a one-dimensional function, we test the approach in a classical sense for the kernel sum and first and second derivatives of a one-dimensional function, using both the Gaussian and Wendland kernel functions, and compare various register sizes against analytical results. Error convergence is exponentially fast in the number of qubits. We extend the method to solve the one-dimensional advection and diffusion partial differential equations, which are commonly encountered in fluids simulations. This work provides a foundation for a more general SPH algorithm, eventually leading to highly efficient simulations of complex engineering problems on gate-based quantum computers. 
\end{abstract}


\end{frontmatter}


\section{Introduction}
\label{sec:Introduction}

Interest in quantum computing and its practical uses has grown dramatically in recent years, as exemplified by Google's claim of `quantum supremacy' \cite{Arute19} and a potential `goldrush' for industry \cite{Gibney19,IBM22,MacQuarrie20}. Quantum computers promise a way to perform highly complicated calculations that are infeasible on classical machines. The power of quantum computation has been well documented \cite{Simon97,Abbas21,Postler22} in many areas such as chemical and materials science \cite{Foulkes01,Bauer20,Reiher17}, high-energy physics \cite{Nachman21,Martinez16,Guan21}, post-quantum cryptography \cite{Bernstein17,Joseph22}, and optimization problems across industry \cite{Yarkoni22,Orus19}. The original motivation for quantum computing was based on Feynman's arguments \cite{Feynman82}: since the world is fundamentally quantum mechanical, it makes sense to use `quantum' machines to simulate both quantum mechanical and classical physical systems \cite{Alexeev21}. 

There already exists a rich ecosystem of quantum numerical algorithms which have applications in modeling, simulation and numerical analysis. A well known example is the Harrow-Hassidim-Lloyd (HHL) algorithm \cite{Harrow09} for solving linear systems of equations to approximate the solution vector. Other prominent methods include quantum walks \cite{VenegasAndraca12,Kendon06,Kadian21}, quantum annealing \cite{Albash18,Bharti22,Chancellor17}, and hybrid quantum-classical algorithms \cite{Zhou20,Cerezo21}. The present era of quantum computing offers new opportunities for numerically modeling physical systems that have real-world applications. Whether quantum methods can reduce the cost of computational fluid dynamics (CFD) simulations is an especially pertinent question for industry \cite{NASA14,NASA22}. Our work joins the growing number of studies on quantum simulations for solving CFD problems. For example, the methods investigated so far include the quantum amplitude estimation algorithm to solve the discretized Navier-Stokes equation \cite{Gaitan20,Gaitan21}; standard form encoding combined with quantum walks to simulate a lattice Boltzmann approach \cite{Budinski22}; quantum Fourier transform to implement vortex-in-cell methods \cite{Steijl18,Steijl19,Steijl20}; linearization methods to simplify nonlinear terms \cite{Liu21,Lloyd20}; and modular quantum circuits to solve the Poisson equation \cite{Cao13,Wang20}.

Our work focuses on the smoothed particle hydrodynamics (SPH) method \cite{Lucy77,Monaghan05}. Mathematically, SPH is an interpolation method that uses a set of disordered points (particles) to express a function in terms of its values at these points. The integral interpolant of any function $\mathcal{A}(\mathbf{r})$ can be expressed as an integral 
\begin{equation}\label{eq:sph-int}
\mathcal{A}(\mathbf{r}) = \int_\Gamma \mathcal{A}(\mathbf{r}')W(\mathbf{r}-\mathbf{r}',h) d\mathbf{r}'
\end{equation}
over the entire space $\Gamma$ for any point $\mathbf{r}$ in space and a smoothing kernel $W$ with smoothing length $h$. Smoothing kernels are often chosen to have a compact support, so the space $\Gamma$ reduces to the support radius of the kernel, often $2h$. The integral interpolant can be approximated by a summation interpolant,
\begin{equation}\label{eq:sph-sum}
\mathcal{A}(\mathbf{r}) = \sum_k m_k \frac{\mathcal{A}(\mathbf{r}_k)}{\rho_k} W(\mathbf{r}-\mathbf{r}_k, h)
\end{equation}
that sums over the set of particles $\{k\}$. Each particle $k$ has mass $m_k$, density $\rho_k$ and velocity $\mathbf{v}_k$ at position $\mathbf{r}_k$. This means a differentiable interpolant of a function can be constructed from its values at the particle level (interpolation points) by using a differentiable smoothing function $W$ \cite{Monaghan05}.

In SPH, the sum (eq. \ref{eq:sph-sum}) is a discrete approximation to the convolution of $\mathcal{A}$ with the Dirac $\delta$-distribution, $\mathcal{A}(\mathbf{r}) = \int \mathcal{A}(\mathbf{r}') \delta(\mathbf{r}-\mathbf{r}')d\mathbf{r}'$ with the kernel $W$ providing a smoothed approximation to the Dirac delta function, $\delta(\mathbf{r})$.

\begin{figure}[ht!]
\centering
\includegraphics[width=\linewidth]{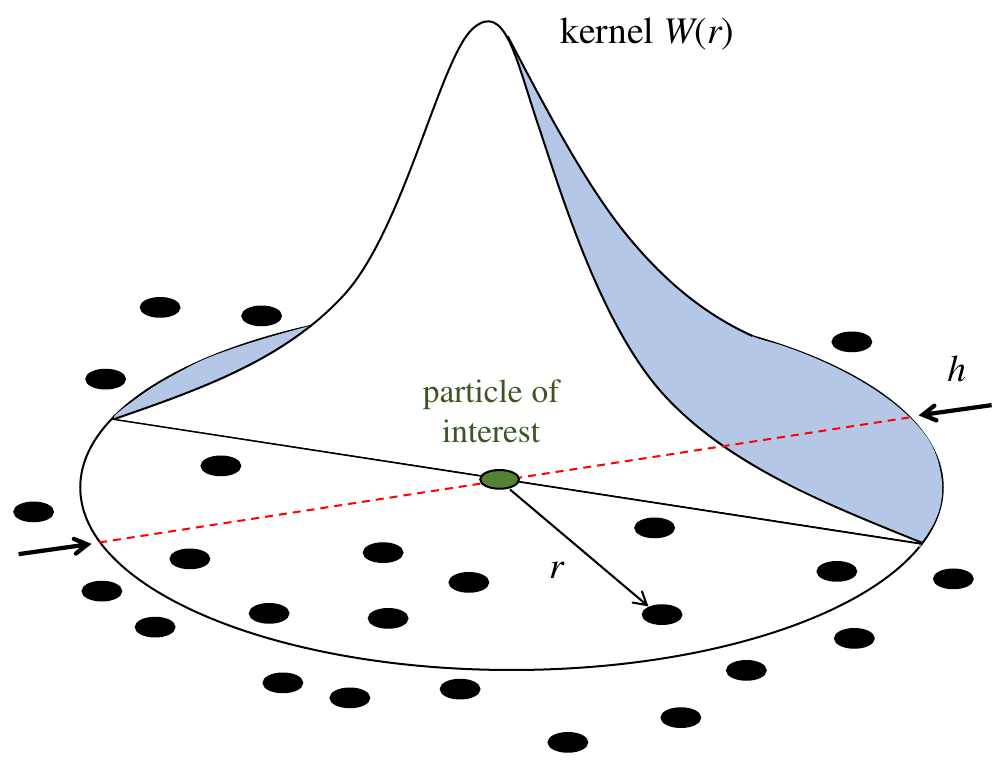}
\caption{Example of a kernel function $W(r)$, with smoothing length $h$. [Adapted from: \href{https://abaqus-docs.mit.edu/2017/English/SIMACAEANLRefMap/simaanl-c-sphanalysis.htm}{Abaqus docs}.]}
\label{fig:kernel}
\end{figure}

Qualitatively, kernels tend to resemble Gaussian profiles but are often constructed to have a compact support (e.g., Fig.~\ref{fig:kernel}) controlled by their smoothing length $h$. This controls the amount of smoothing and hence how strongly the value of $\mathcal{A}$ at position $\mathbf{r}$ is influenced by the values in its proximity. The smoothing effect increases with $h$. Other key qualities for $W$ include symmetry, positivity, normalization, and convergence to a Dirac $\delta$-function in the limit $h\to0$ \cite{Monaghan05,Price12}. It is essential to satisfy the normalization and Dirac $\delta$ conditions to ensure that the approximation to (eq.~\ref{eq:sph-int}) remains valid. To properly discretize second-order partial differential equations, the kernel should be at least twice continuously differentiable.

\begin{figure}[ht!]
\centering
\begin{subfigure}[]{\linewidth}
\centering
\includegraphics[width=\linewidth]{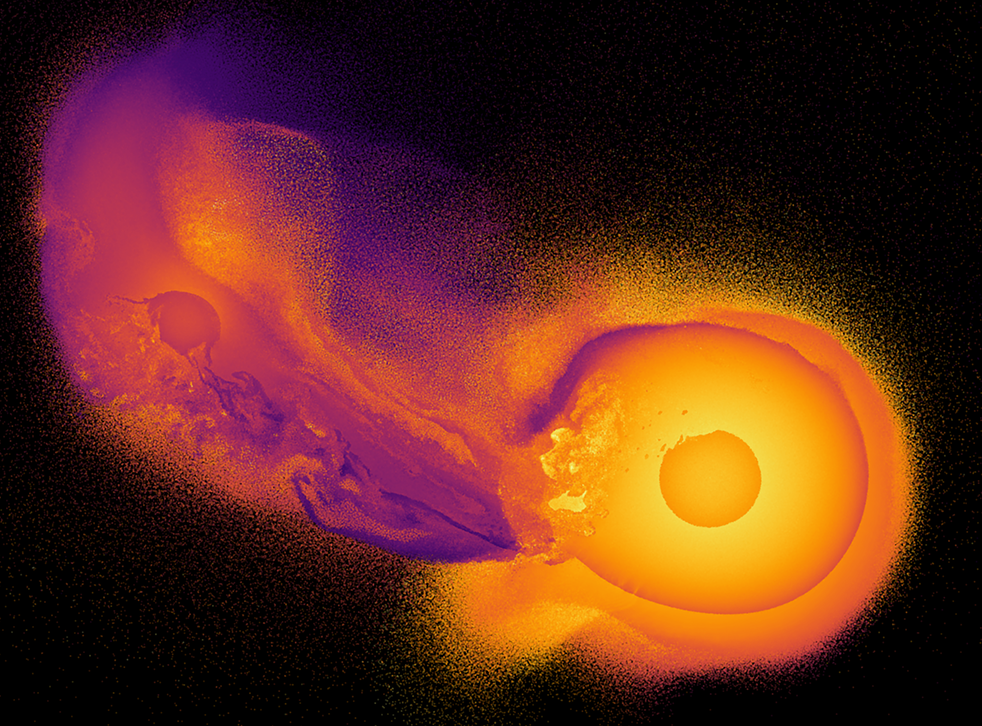}
\caption{Mid-collision snapshot of colliding planets using $10^8$ SPH particles, colored by their material and internal energy \cite{Kegerreis19}.}
\label{fig:sph_astro}
\end{subfigure}
\hfill
\begin{subfigure}[]{\linewidth}
\centering
\includegraphics[trim = 35mm 0mm 0mm 0mm, clip, width=\linewidth]{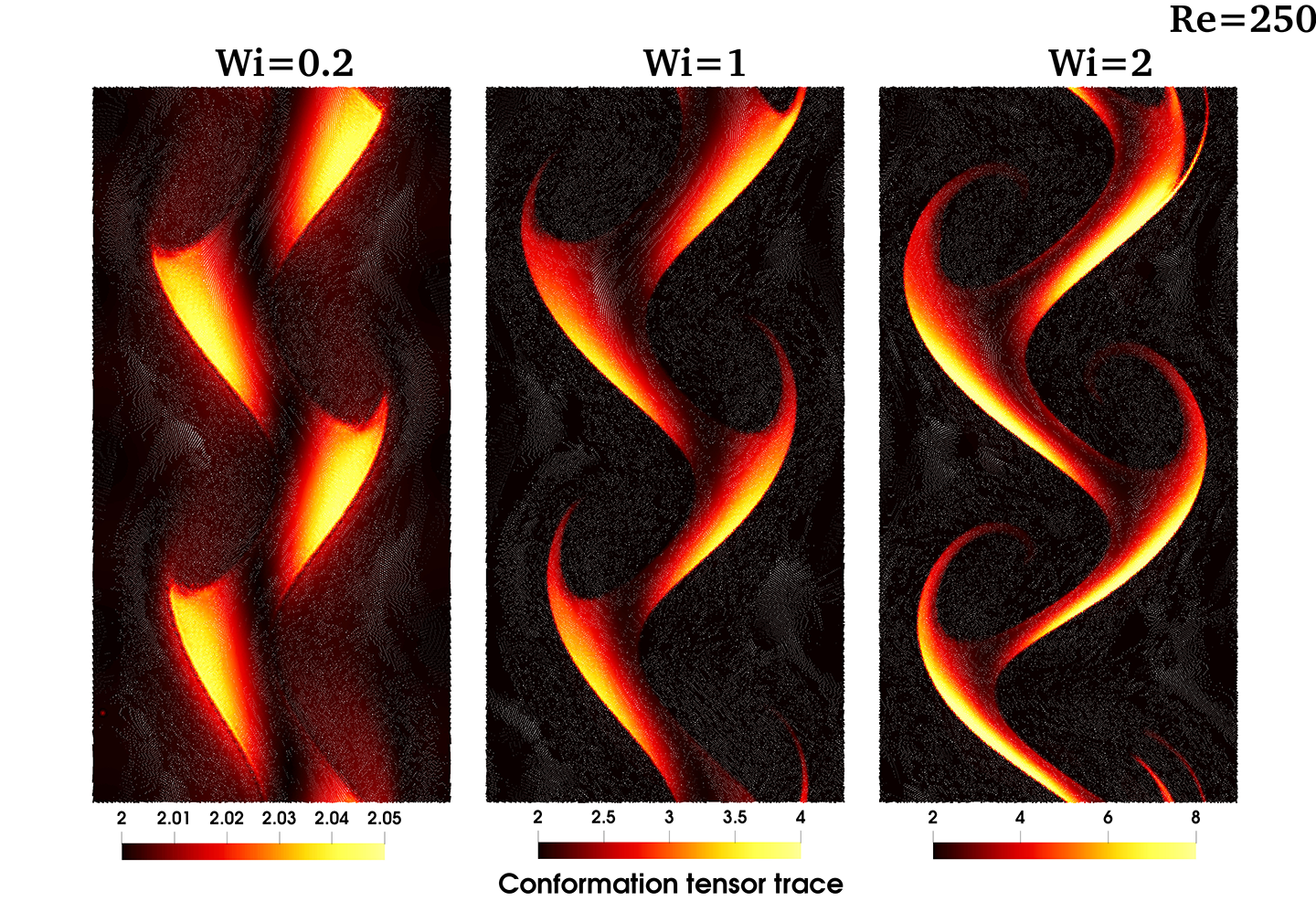}
\caption{Simulations of a Kelvin-Helmholtz instability of a viscoelastic jet in contraflow with a (nearly) Newtonian fluid. The color shows viscoelastic conformation tensor trace, indicating molecular deformation. Using 115,200 particles, increasing elasticity ($Wi$) causes an increased growth rate, finer filaments and increased transverse mixing. Credit: Jack King and Steven Lind, University of Manchester \href{https://fluids.ac.uk/gallery/zoomify/744}{[link]}}
\label{fig:kh_instability}
\end{subfigure}
\caption{Examples of SPH in astrophysics and engineering.}
\label{fig:sph_examples}
\end{figure}

As a flow solver, SPH is typically a Lagrangian method that uses particle interpolation to approximate continuous field variables. These particles carry the system's physical properties (such as mass and temperature) and we can construct the governing equations of the discrete system to conserve mass, energy and momentum. The Lagrangian nature gives clear advantages over traditional mesh-based Eulerian methods. For example it does not suffer from mesh distortions that affect the numerical accuracy and stability when simulating large material deformations. This is useful, for example, in highly compressible flows as the Lagrangian particles naturally resolve the variable density regions. Similarly, SPH can model violently deforming and dynamic interfaces without using special treatments required for meshes (e.g., mesh re-zoning). This has led to its adoption by many application areas such as astrophysics \cite{Springel10,Price12,Kegerreis19} (see example Fig.~\ref{fig:sph_astro} from \cite{Kegerreis19}) and engineering fluid flows \cite{Monaghan12, king2021high}  (see example Fig.~\ref{fig:kh_instability}, based on work of \cite{king2021high}). There are many similar meshless numerical methods based on a force or kernel sum, as in SPH. Examples include the force field calculation in classical molecular dynamics \cite{Monticelli13}, smoothed dissipative particle dynamics (DPD) \cite{Hoogerbrugge92,Groot97}, and general purpose radial basis functions (which often use the Gaussian and Wendland kernel types in our work) \cite{Buhmann03}. The quantum algorithm in our work is just as applicable and readily generalizable to these methods.

Indeed, there may be computational advantages and opportunities for advanced multi-scale applications and coupled approaches. Each SPH particle represents a finite volume in continuum scale. It is similar to the classic molecular dynamics (MD) method \cite{Frenkel01} that uses particles to represent molecules in nano-scale, and the DPD method that uses a particle to represent a small cluster of molecules in mesoscale. Thus, it is natural to generalize or extend SPH to smaller scales, or to couple SPH with molecular dynamics and dissipative particle dynamics for multiple scale applications, especially in biophysics, and biochemistry.

Based on the general interpolation (eq. \ref{eq:sph-sum}), SPH can be used as a general PDE solver: it approximates any differential operators. As a discrete particle method, the SPH system may also be described using the classical Hamiltonian. Hence this provides a natural link with quantum computing and a potential route to efficient and practical quantum nonlinear PDE solvers. Years of theoretical work on quantum simulators have provided efficient quantum algorithms that run in almost linear time \cite{Berry07,Daley22} for calculating the time evolution under a quantum Hamiltonian. Given that the classical Hamiltonian underpins SPH physics, there should be at least one natural mapping of SPH to quantum computers that we can exploit for quantum enhancement. For our work, we take a different approach - we investigate the SPH approximation of a function and solve partial differential equations using quantum subroutines.

In this work our primary focus is on devising quantum subroutines for the SPH discretization. SPH accuracy relies on using a large number of particles in the simulation. Theoretically, quantum machines could allow exponentially more particles to be used without significantly increasing the runtime. For example, every additional qubit in our proposed algorithm would double the number of SPH particles. While we await testbed hardware, how we can achieve this in practice remains unclear. It may be that hardware memory constraints (on either quantum or hybrid classical devices) provide a practical limitation to what we may simulate. Nevertheless, such quantum subroutines raise the possibility of highly resolved simulations, potentially deployed within multi-resolution schemes in sub-domains within a lower-resolution classical simulation. This could offer an efficient route to high resolution and even Direct Numerical Simulation (DNS) of challenging turbulent flows using SPH.

The structure of our paper is as follows. First we introduce the core ideas behind our quantum method such as quantum registers. As an example, we calculate the first and second derivatives of a smooth, well-behaved function, laying out each step of the domain discretization and quantum encoding processes. We numerically solve the SPH approximation of the first and second derivatives for different numbers of qubits and SPH kernels. Note that we simulate the quantum algorithm on a classical machine. Then we adapt the method to solve the advection and diffusion equations, and compare with the results from classical numerical methods. We finish with a discussion on how to improve the method by identifying the bottlenecks and ideas for future work.

\section{Quantum principles}
\label{sec:Quantum_principles}

There are many excellent introductory resources on quantum computing science (see e.g. \cite{Nielsen10,Portugal22}) so we will not present a detailed review here. Instead in this section we briefly introduce the quantum subroutines that we use in our method, namely quantum register and inner product.

\subsection{Quantum registers}
\label{sec:Quantum_registers}

The central idea behind quantum speed‐ups over conventional computing is due to quantum bits (qubits) and hence quantum superposition and entanglement \cite{Horodecki09}. In this work, we focus on gate-based quantum computing models \cite{Nielsen10}. These are the basis of devices by IBM, Rigetti and others. 

In classical computers, we store the intermediate results of a program in an electronic circuit, the \textit{register}. The contents of such a register consist of bits which are changed with each operation. In the gate-based quantum computing model, quantum registers and qubits are respectively the analog of classical registers and bits. Now the quantum mechanical qubit state can represent two complex numbers: a quantum register containing $m$ entangled qubits can represent $2^m$ complex numbers and every quantum operation on the register acts on all superpositions simultaneously \cite{Nielsen10}.

Measuring the registers would output strings of bits like classical computer registers. If each qubit in the register is in a superposition, then the register of $m$ qubits is in a superposition of all $2^m$ possible bit strings that may be represented using $m$ bits. The state space for a quantum register is a linear combination of $m$ basis vectors $\ket{k}$. The superposition state of length $m$ is
\begin{align}
\ket{\psi_m} = \sum_{k=0}^{2^m - 1} \alpha_k \ket{k}.
\end{align} 
For example a three-qubit register would be represented as $\ket{\psi_3} = \alpha_0 \ket{000} + \alpha_1 \ket{001} + \alpha_2 \ket{010} + \alpha_3 \ket{011} + \alpha_4 \ket{100} + \alpha_5 \ket{101} + \alpha_6 \ket{110} + \alpha_7 \ket{111}$ with complex numbers $\alpha_k$. The probability of observing a particular bit string upon measuring the register is $\vert\alpha_k\vert^2$, and the quantum register must satisfy the normalization condition
\begin{align}\label{eqn:sum_of_probabilities}
\sum_{k=0}^{2^m - 1} |\alpha_k|^2 = 1.
\end{align}

\subsection{Inner products}

Multiplying two vectors together using the inner product to produce a scalar is a useful operation in many numerical algorithms. For example, matrix multiplication can be broken down into successive applications of the inner products of rows and columns to form the entries in the resultant matrix. Estimating, rather than evaluating, inner products is also important especially when considering large vectors. It is used in HHL \cite{Harrow09} and quantum machine learning algorithms \cite{Lloyd13,Rebentrost14,Cai15} and has attracted renewed interest as a fundamental primitive. The inner product is core to our proposed SPH algorithm, and in Section \ref{sec:Discussion} we review some interesting existing quantum algorithms for inner products that may be used as supporting subroutines in the future.

\section{SPH using a quantum register}
\label{sec:SPH_quantum_register}

In this section, we present two examples. We start by finding the SPH approximation of a one-dimensional function $f(x)$ and its first and second derivatives. Extending this procedure, we solve the one-dimensional advection and diffusion equations.

\subsection{Calculating derivatives}
\label{sec:derivatives}

\begin{figure*}[ht!]
\centering
\includegraphics{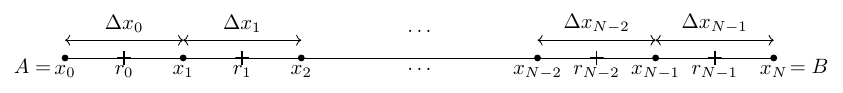}
\caption{Domain discretization with particle locations $r_k$ and sizes $\Delta x_k$.}
\label{fig:domain_discretization}
\end{figure*}

We define a one-dimensional function $f(x)$ on the finite domain $x \in [A,B]$ where $A < B$ are constants, using an $m$ qubit quantum register. Let $\{ x_j \}$ be a partition of $[A,B]$ such that
\begin{align}\label{eq:intervals}
A = x_0 < x_1 < \ldots < x_j < \ldots < x_{N - 1} < x_{N} = B,
\end{align}
where $N=2^m$ is the number of subintervals. Each $x_j$, where $j \in \{0,1,\ldots, N \}$, defines the edge of a subinterval. The width of the $k${th} subinterval is
\begin{align}
\Delta x_k = x_{k+1} - x_k, \hspace{0.5cm} k \in \{0,1, \ldots, N-1 \}.
\end{align}
Each SPH particle is located at the centre of the respective subinterval so that the particle locations are given by 
\begin{align}
r_k = \frac{x_{k+1} + x_k}{2}, \hspace{0.5cm} k \in \{0,1, \ldots, N - 1 \}.
\end{align}
The domain discretization is shown in Fig. \ref{fig:domain_discretization}. The function value at each particle location is $f_k = f(r_k)$.

The one-dimensional SPH approximation of any function is 
\begin{equation}\label{eqn:SPH_approximation}
f(r) \approx \sum_k f_k \Delta x_k W(r - r_k, h) 
\end{equation}
where $W(r,h)$ is a known kernel function. Derivatives of $f$ can then be easily estimated by replacing the kernel with the required derivative, as in SPH the derivative is found by taking an exact derivative of $W$ in approximation (eq. \ref{eqn:SPH_approximation}) \cite{Monaghan05}. We wish to evaluate this SPH approximation using a quantum computer therefore it is necessary to encode the values in a quantum register. 

First we rewrite the summation (eq. \ref{eqn:SPH_approximation}) as an inner product of two vectors $f \approx \mathbf{a} \cdot \mathbf{W}$ where 
\begin{gather}
\mathbf{a} = [f_0 \Delta x_0, f_1 \Delta x_1, \ldots, f_{N-1} \Delta x_{N-1}], \nonumber \\
\mathbf{W} = [W_{r,0}, W_{r,1}, \ldots, W_{r,N-1}]
\end{gather}
and $W_{r,k} = W(r-r_k,h)$. Initially, for simplicity, we assume that all subintervals are equal hence $\Delta x_1 = \Delta x_2 = ... = \Delta x$ (Fig. \ref{fig:domain_discretization}). By effectively fixing the SPH particle positions, we negate an important (Lagrangian) element of SPH. However we only do this to pare back the mathematical details and show the mechanisms behind the quantum algorithm more clearly. If the SPH particles could move freely, we would need another quantum register(s) to keep track of the particle positions (and other properties). Ipso facto this is an area for further investigation.

Next we encode the vectors in a quantum register by calculating appropriate normalization factors and augmenting the entries with complex values.

For the vector $\mathbf{a}$ we define the quantum state 
\begin{align}\label{eq:a_ket}
\ket{a} = \frac{\mathbf{a}}{\lVert \mathbf{a} \rVert},
\end{align}
where $\lVert \cdot \rVert$ denotes the Euclidean (L2) norm. If we have a large number of SPH particles then calculating this norm directly is computationally expensive and defeats the objective of encoding the values in a quantum register. However we use an approximation of the form
\begin{align}\label{eqn:norm_approximation}
\lVert \mathbf{a} \rVert \approx 
\frac{1}{\sqrt{N}} 
\left(
\int_{A}^{B} 
|f|^2 dx
\right)^{1/2}
\end{align}
using the function $f$ or a smaller number of its values. As $N$ becomes large this approximation becomes increasingly accurate. Hence we rewrite $\mathbf{a}$ using an $m$ qubit quantum register as $\lVert \mathbf{a} \rVert \ket{a}$. The norm of $\ket{a}$ is unity so this is a legitimate quantum state. 

Encoding the kernel vector $\mathbf{W}$ requires a little more ingenuity since calculating the Euclidean norm of this vector is computationally expensive. 

\begin{enumerate}

\item First we scale the vector using $\nu=\max(|W(r,h)|)$ so that the maximum/minimum value is $\pm 1$. We define the scaled vector $\widetilde{\mathbf{W}} = \mathbf{W} / \nu$. If $W(r,h)$ is a symmetric kernel function then $\nu=W(0,h)$. However $\nu$ will vary for different kernels and their respective derivatives. 

\item We scale the vector again using the number of subintervals $N$ to give $\widehat{\mathbf{W}} = \mathbf{W} / (\nu N)$ so that the largest absolute value of $\widehat{\mathbf{W}}$ is $1/N$. 

\item We create a quantum state using the values in vector $\widehat{\mathbf{W}}$ plus a complex term which we choose to satisfy the normalization conditions of a quantum register (eq. \ref{eqn:sum_of_probabilities}).

\end{enumerate}

Suppose we add an imaginary part $b_{r,k}$ to each value in $\widehat{\mathbf{W}}$ to create a quantum state of the form
\begin{align}\label{eqn:kernel_quantum_state}
\ket{W} = 
\begin{bmatrix}
\widehat{W}_{r,0} + i b_{r,0} \\
\widehat{W}_{r,1} + i b_{r,1} \\
\vdots \\
\widehat{W}_{r,N-1} + i b_{r,N-1}
\end{bmatrix}.
\end{align}
If we choose the $b_{r,k}$ values appropriately then $\ket{W}$ is a legitimate quantum state. If 
\begin{align}
b_{r,k} = \sqrt{\frac{1}{N} - \widehat{W}_{r,k}^2 }
\end{align}
then
\begin{align}
\left| \widehat{W}_{r,k} + i b_{r,k} \right|^2 = \widehat{W}_{r,k}^2 + b_{r,k}^2 = \frac{1}{N}.
\end{align}
Hence
\begin{align}
\sum_{k=0}^{N-1} \left| \widehat{W}_{r,k} + i b_{r,k} \right|^2 = \sum_{k=0}^{N-1} \frac{1}{N} = 1,
\end{align}
as required. The kernel function values are encoded in a quantum state, albeit with additional imaginary parts. Both approaches for encoding $\ket{a}$ and $\ket{W}$ provide legitimate quantum states. Depending on the variable being encoded, one approach may be favored over the other. For example, our $\ket{a}$ is designed to contain a physical flow variable. While we can directly calculate the L2 norm, in this context the norm may be related (approximately or otherwise) to global, perhaps constant, flow measures that are already known or readily available. If $\ket{a}$ encodes velocity data, then the L2 norm scales to kinetic energy. Similarly, if $\ket{a}$ encodes density then $\lVert \mathbf{a} \rVert$ is related to a constant global measure of mass. Being able to use physical arguments makes the encoding and computation more efficient. On the other hand, normalizing via an additional complex term $b_{r,k}$ provides a valid state more efficiently if no norm approximations are available. Rather than taking direct and repeated arithmetic operations to find the norm, we can construct a state using $b_{r,k}$ values. These ultimately take no part in the calculation since we only require the real part of the inner product.

We must now use the quantum states $\ket{a}$ and $\ket{W}$ to reconstruct the SPH approximation of our function $f$. In general, encoding the kernel is likely to be expensive. Depending on the choice of kernel, this may involve exponentiation operations or conditional statements (e.g. for piecewise kernels). It is also unknown whether there will be sufficient memory on near-term quantum devices to retain variable values, as processor speed is likely to prioritized over memory (or QRAM). Hence given the memory limitations and the need to repeat calculations on quantum hardware, even modest computational savings via different encoding options are welcome. It should also be noted that if both states are normalized by adding an imaginary part then spurious real elements will be created from the combination of the two imaginary parts. In this respect, using two different normalization approaches together has further benefit.

Taking the inner product of $\ket{a}$ and $\ket{W}$ gives 
\begin{align}
\langle a \vert W \rangle = &\frac{1}{\lVert \mathbf{a} \rVert} \bigg[ f_0 \Delta x_0 ( \widehat{W}_{r,0} + i b_{r,0}) \nonumber\\
&+  f_1 \Delta x_1 ( \widehat{W}_{r,1} + i b_{r,1}) \nonumber\\
&+ \ldots \nonumber\\
&+ f_{N-1} \Delta x_{N-1} ( \widehat{W}_{r,N-1} + i b_{r,N-1} ) \bigg].
\end{align}
As noted, the imaginary part is not required when $\ket{a}$ is purely real (which it is in all examples in our paper). However due to physical arguments, we must include it so that $\ket{W}$ is a valid state.

Therefore multiplying through by $\nu N \lVert \mathbf{a} \rVert$ we have 
\begin{align}
\nu N \lVert \mathbf{a} \rVert \langle a \vert W \rangle = \sum_{k=0}^{N-1} f_k \Delta x_k \left( W_{r,k} + i \nu N b_{r,k} \right).
\end{align} 

Retaining only the real part of the inner product,
\begin{align}
f(r) \approx \sum_{k=0}^{N-1} f_k \Delta x_k W_{r,k} = \nu N \lVert \mathbf{a} \rVert \Re \langle a \vert W \rangle.
\end{align}
This is equivalent to the SPH approximation of a function but calculated using a quantum register.

An $m$ qubit register, storing $N=2^m$ values, can be used to perform the SPH approximation on a quantum register. Subsequently, the swap test or another method (see section \ref{sec:Discussion}) determines the inner product $\langle a \vert W \rangle$. Its efficiency relies on the values $\nu$ and $\lVert \mathbf{a} \rVert$ being known and there being a fast method to encode the quantum states $\ket{a}$ and $\ket{W}$. The kernel function and its encoding can easily be replaced by equivalent derivative kernels so that derivatives of the function may be approximated, provided that $\nu$ is altered accordingly.  

In this section, we develop a method for encoding the SPH approximation of a function in a quantum register. The quantum computation required for this procedure can be simulated on a classical machine when there are fewer than 10 qubits. To compare, some current real devices contain 50-100 qubits. Simulating over roughly 40 qubits with a classical computer is beyond the reach of current HPC. Now we show simulations of the one-dimensional SPH approximation of a function and its derivatives on a quantum computer for various register sizes and kernel functions. 

We test the scheme by approximating the scaled "Witch of Agnesi" function
\begin{align}
\label{eqn:Runge_function}
f(x) = \frac{1}{1 + 25 x^2}
\end{align} 
and its first 
\begin{equation}
f'(x) = -\frac{50x}{(1 + 25x^2)^2}
\end{equation}
and second derivatives 
\begin{equation}
f''(x) = 50 \frac{75 x^2 - 1}{(1 + 25x^2)^3}
\end{equation}
for different register sizes, using both the Gaussian
\begin{align}
\label{eqn:Gaussian_kernel}
W(r,h) = \frac{e^{-q^2}}{\sqrt{\pi} h} ,
\end{align}
and Wendland kernels,
\begin{align}
\label{eqn:Wendland_kernel}
W(r,h) = 
\begin{cases}
\frac{3}{4h} \left(1 - \frac{1}{2} q \right)^4 \left( 2q + 1 \right), & 0 \leq q \leq 2, \\
0, & q > 2,
\end{cases}
\end{align}
where $q = |r| / h$. A key element of the quantum numerical scheme outlined above is the constant $\nu=\max(|W(r,h)|)$ which is used to scale the weight vector and is different for each kernel. In Table \ref{table:nu_values}, we show $\nu$ for both kernels and their first and second derivatives. 

\begin{table}[ht!]
\centering
\begin{tabular}{|c|c|c|}
\hline
Kernel & Derivative & $\nu$ \\
\hline 
\multirow{3}{*}{Gaussian} & - & $1/(\sqrt{\pi} h)$ \\
 & 1st & $\sqrt{2}e^{-1/2} / (\sqrt{\pi} h^2)$ \\
 & 2nd & $2/(\sqrt{\pi}h^3)$ \\ \hline
\multirow{3}{*}{Wendland} & - & $3/(4h)$ \\
 & 1st & $405/(512h^2)$ \\
 & 2nd & $15 / (4 h^3)$ \\
\hline
\end{tabular}
\caption{Values of $\nu=\max(|W(r,h)|)$ for the Gaussian and Wendland kernels and their first and second derivatives.}
\label{table:nu_values}
\end{table}

\begin{figure}[ht!]
\centering
\includegraphics[width=\linewidth]{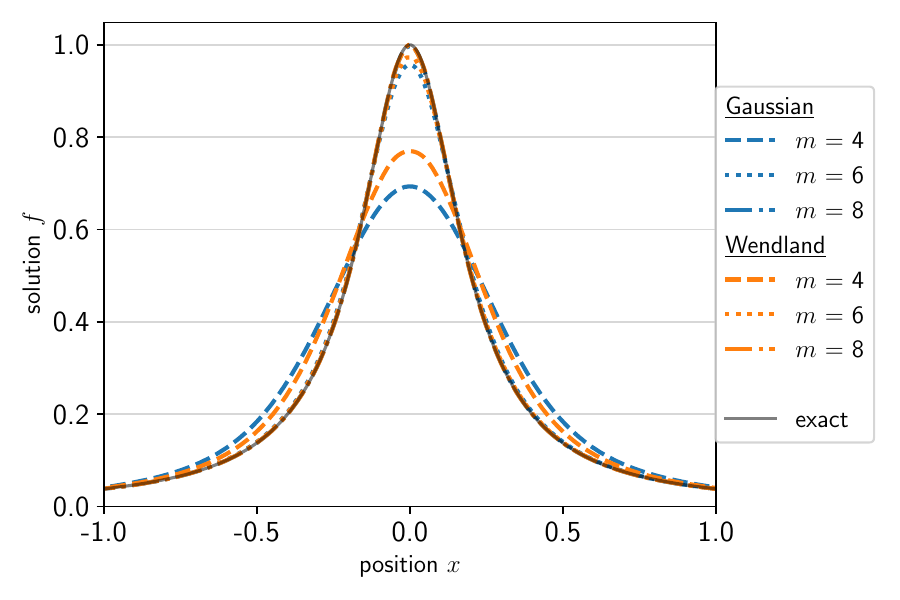}
\caption{The function \eqref{eqn:Runge_function} and its quantum SPH approximations for $m=4,6,8$ qubits.}
\label{fig:function}
\end{figure}

\begin{figure}[ht!]
\centering
\includegraphics[width=\linewidth]{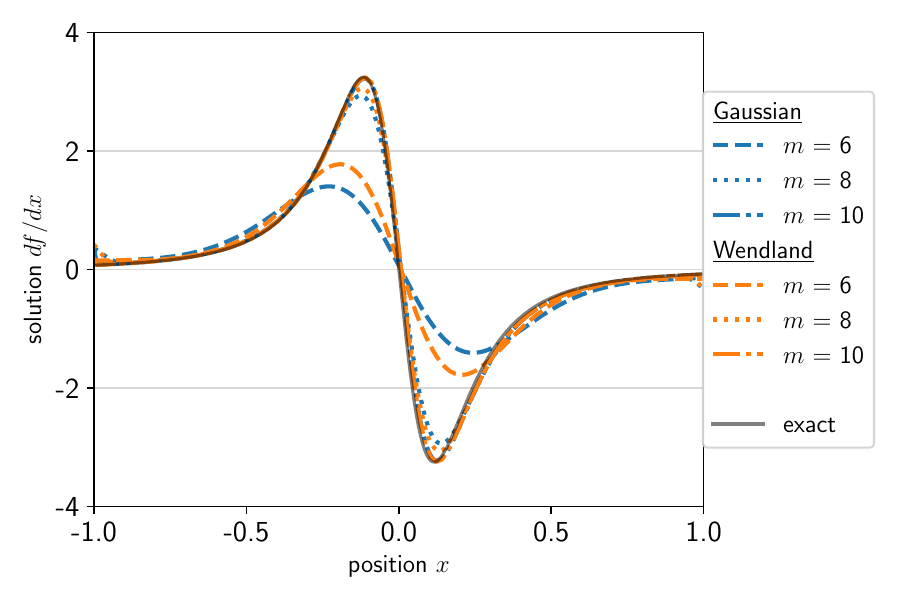}
\caption{The first derivative function and its quantum SPH approximations for $m=6,8,10$ qubits.}
\label{fig:deriv1}
\end{figure}

\begin{figure}[ht!]
\includegraphics[width=\linewidth]{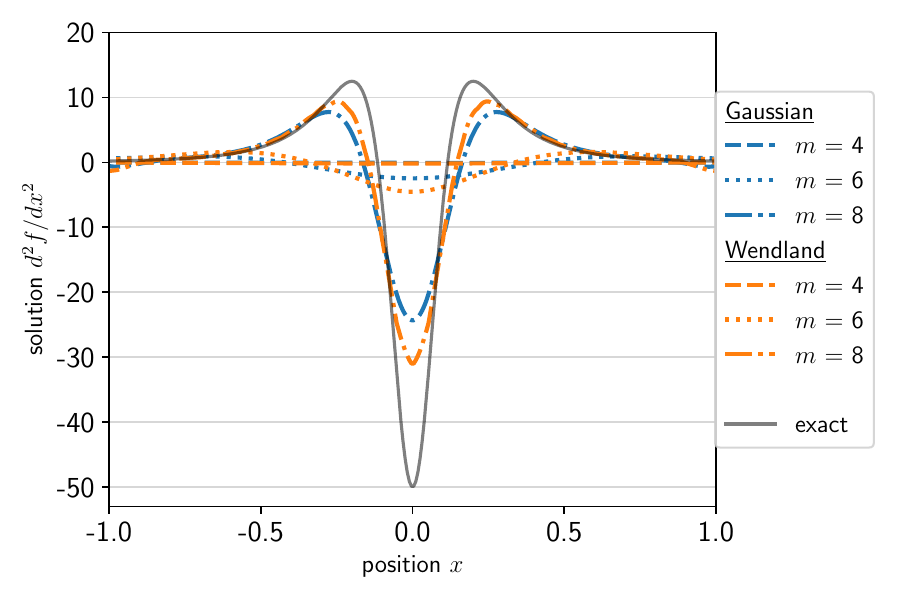}
\caption{The second derivative function and its quantum SPH approximations for $m=4,6,8$ qubits.}
\label{fig:deriv2}
\end{figure}

\begin{figure}[ht!]
\centering
\includegraphics[width=\linewidth]{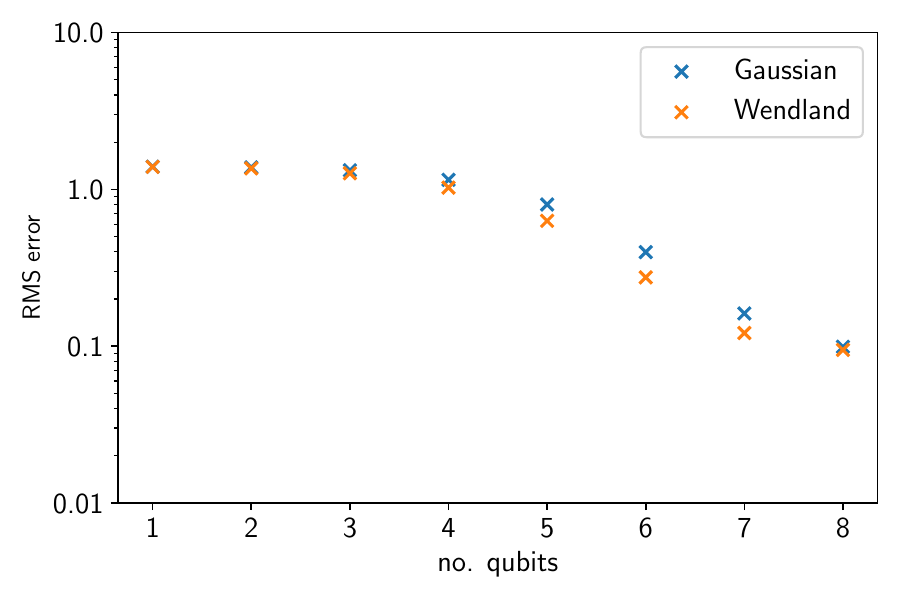}
\caption{RMS error for Gaussian and Wendland kernel sum approximations of \eqref{eqn:Runge_function} for various $m$ qubit registers.}
\label{fig:rms_error}
\end{figure}

Figure \ref{fig:function} shows the quantum SPH approximation of the function (eq. \ref{eqn:Runge_function}), using various register sizes $m$, for the Gaussian and Wendland kernels. For each $m$-qubit approximation, we use smoothing length $h=4/2^m=2\Delta x$; $2^m$ particles in the domain and four additional boundary particles to complete kernel support at each edge of the domain. We measure the function approximations at $n=300$ uniformly distributed points $x'_j$ in the domain $x \in [-1,1]$. Note, these are not the domain discretization points, but simply points in the domain where we calculate the approximation. Figure \ref{fig:deriv1} shows approximations of the first derivative of (eq. \ref{eqn:Runge_function}) for the Gaussian and Wendland kernels. Figure \ref{fig:deriv2} shows the approximations of the second derivative of (eq. \ref{eqn:Runge_function}). Figure \ref{fig:rms_error} shows the root-mean-square (RMS) error 
\begin{align}\label{eq:rms}
\text{RMS} = \sqrt{\frac{\sum_{j=1}^n \left(f(x'_j) - f_j \right)^2 }{n}},
\end{align} 
where $f(x'_j)$ is the exact value of $f(x)$ at the point $x'_j$ and $f_j$ is the quantum SPH approximation, as a function of the register size $m$ for both the Gaussian and Wendland kernel approximations of the function \eqref{eqn:Runge_function}. 

In classical SPH simulations, this is equivalent to calculating eq. (\ref{eqn:SPH_approximation}) for function $f(r_j)$. The numerical bottlenecks occur when finding the pair-wise particle separations $r_j-r_k$ and using these values to evaluate the summation. There also exists various neighbor searching algorithms of varying efficiency. We can store details of these distances in efficient data structures but searching and summation must generally be repeated for each SPH particle $j$ and at each timestep (if applicable). In comparison, the main bottlenecks in the quantum method occur when encoding the SPH operators and domain discretization in a quantum register, performing the inner product for each timestep, and quantum readout steps. We note that the quantum method presented here is a simplification that uses SPH particles fixed in space. In a general setting with freely moving particles, we would need to devise an efficient neighbor search subroutine to address the potential bottlenecks.

\subsection{Solving the advection equation}

The advection equation is a fundamental partial differential equation describing the transport of some physical quantity. In the one-dimensional case, the advected quantity $u(x,t)$ changes in space $x$ and time $t$ according to the partial differential equation
\begin{equation}\label{eq:advection}
\frac{\partial u(x,t)}{\partial t} + c(x,t) \frac{\partial u(x,t)}{\partial x} = 0
\end{equation}
for advection velocity $c(x,t)$ \cite{Liu03}. 

For simplicity, we consider the linear advection equation with constant velocity $c(x,t)=c$. Hence the initial condition gives solutions that are uniform translations of the initial profile, $u(x,t=0)=u_0(x-ct)$.

Since any quantity $f$ can be written as eq. (\ref{eqn:SPH_approximation}), its spatial derivative is
\begin{equation}
\frac{\partial f(x)}{\partial x} \approx \sum_{j=1}^N f(x_j) \Delta x_j \nabla W(x-x_j,h).
\end{equation}
The following form is commonly used in the SPH community
\begin{equation}\label{eq:dudx}
\frac{\partial u}{\partial x} \approx \sum_{j=1}^N (u_j-u_i) \Delta x_j \nabla_i W(x_i-x_j,h)
\end{equation}
because it is more accurate by being zeroth-order consistent. We use the Courant-Friedrichs-Lewy (CFL) condition to define the timestep size
\begin{equation}\label{eq:cfl-advection}
\Delta t \leq \frac{\Delta x}{c}.
\end{equation}
Then we express the advection equation in SPH form as
\begin{equation}\label{eq:advection_sph}
u_i^{(n+1)} = 
u_i^{(n)} - c\Delta t \sum_{j=1}^N 
(u_j^{(n)}-u_i^{(n)}) \Delta x_j \nabla_i W_{ij}(h)
\end{equation}
where the $(u_j^{(n)}-u_i^{(n)})$ terms are analogous to $f_k$ in eq. (\ref{eqn:SPH_approximation}) and $W_{ij}(h)=W(x_i-x_j,h)$. 

We follow a similar procedure as in the previous example to encode the function into a quantum register. We rewrite the summation as an inner product $\mathbf{a} \cdot (\nabla_i\mathbf{W})$ with vectors
\begin{equation}
\mathbf{a} = 
\begin{bmatrix}
(u_{1}^{(n)}-u_i^{(n)}) \Delta x_{1} \\
(u_{2}^{(n)}-u_i^{(n)}) \Delta x_{2} \\
\vdots \\
(u_{N}^{(n)}-u_i^{(n)}) \Delta x_{N} 
\end{bmatrix},
\quad
\mathbf{W} = 
\begin{bmatrix}
W_{i,1} \\ W_{i,2} \\ \vdots \\ W_{i,N}
\end{bmatrix}
\end{equation}
at time $t_n$. When defining the quantum state $\vert a \rangle = \mathbf{a}/\norm{\mathbf{a}}$, we use an approximation to efficiently calculate
\begin{equation}
\norm{\mathbf{a}} \approx \frac{1}{\sqrt{N}} 
\left( 
\int_A^B \vert u^{(n)}-u_i^{(n)} \vert^2 dx
\right)^{1/2}.
\end{equation}

Since $\nabla$ is a linear operator, it is trivial to calculate $\vert \nabla_i W \rangle$ from $\nabla_i \mathbf{W}$. We scale vector $\nabla_i \widehat{\mathbf{W}} = \nabla_i \mathbf{W}/(\nu N)$ where $\nu=\max(\vert \nabla W(r,h) \vert)$. Therefore we may write the real part of the inner product
\begin{equation}
\Re \langle a \vert \nabla_r W \rangle = \frac{1}{\nu N \norm{\mathbf{a}}} \sum_{j=1}^{N} (u_j^{(n)}-u_i^{(n)}) \Delta x_j \nabla_i W_{i,j}
\end{equation}
and
\begin{equation}\label{eq:advection_sph0}
u_i^{(n+1)} = u_i^{(n)} - c\Delta t \; \nu N \norm{\mathbf{a}} \Re \langle a \vert \nabla_i W \rangle.
\end{equation}

\begin{table*}[ht!]
\centering\small
\begin{tabular}{c|ccccccc|cc|}
 & initial & & & & & smoothing & CFL & \multicolumn{2}{c|}{} \\
 & Gaussian & $c$ & $t_f$ & $x\in[A,B]$ & $m$ qubits & length & condition & \multicolumn{2}{c|}{RMS (4 dp)} \\
 & $[\beta,L]$ & & & & & $h$ & $c\frac{\Delta t}{\Delta x}$ & \multicolumn{2}{c|}{} \\ \hline 
 & & & & & & & & Gaussian & 0.0105 \\
Fig. \ref{fig:advection1} & $[0,0.4]$ & 1 & 0.5 & $[-2,2]$ & 8 & $16/2^m$ & 0.1 & Wendland & 0.0061 \\ 
 & & & & & & & & Lax-Wendroff & 0.0095 \\ \cline{9-10}
 & & & & & & & & Gaussian & 0.0033 \\
 & $[0,0.4]$ & 1 & 0.5 & $[-2,2]$ & 8 & $8/2^m$ & 0.1 & Wendland & 0.0019 \\ 
 & & & & & & & & Lax-Wendroff & 0.0100 \\ \hline
 & & & & & & & & Gaussian & 0.0003 \\
Fig. \ref{fig:advection2} & $[0.15,0.7]$ & 1.3 & 0.5 & $[-4,4]$ & 10 & $16/2^m$ & 0.1 & Wendland & 0.0010 \\ 
 & & & & & & & & Lax-Wendroff & 0.0028 \\ \cline{9-10}
 & & & & & & & & Gaussian & 0.0041 \\
 & $[0.15,0.7]$ & 1.3 & 0.5 & $[-4,4]$ & 10 & $16/2^m$ & 1 & Wendland & 0.0050 \\ 
 & & & & & & & & Lax-Wendroff & 0.0038 \\ \hline
\end{tabular}
\caption{Data table for advection equation simulation.}
\label{tab:advection-table}
\end{table*}

\begin{figure*}[ht!]
\centering
    \begin{subfigure}{0.45\textwidth}
    \centering
    \includegraphics[width=\linewidth]{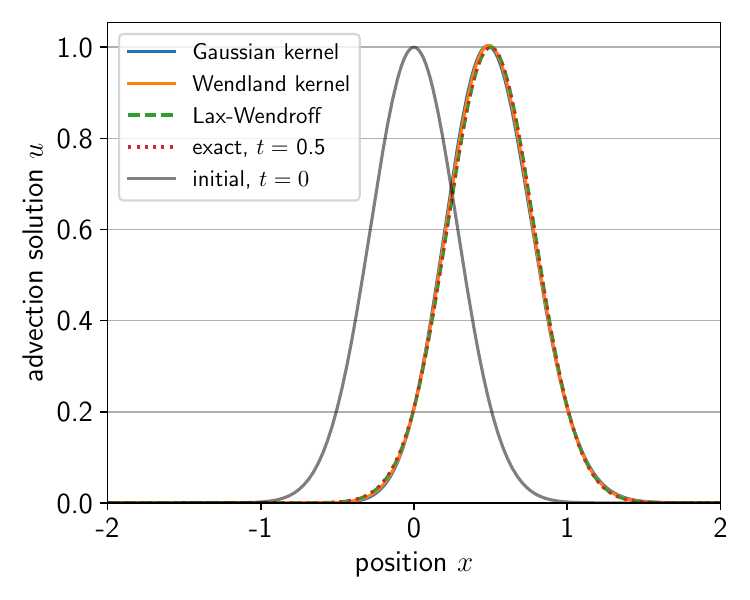}
    \caption{}
    \label{fig:advection1}
    \end{subfigure}
\hfill
    \begin{subfigure}{0.45\textwidth}
    \centering
    \includegraphics[width=\linewidth]{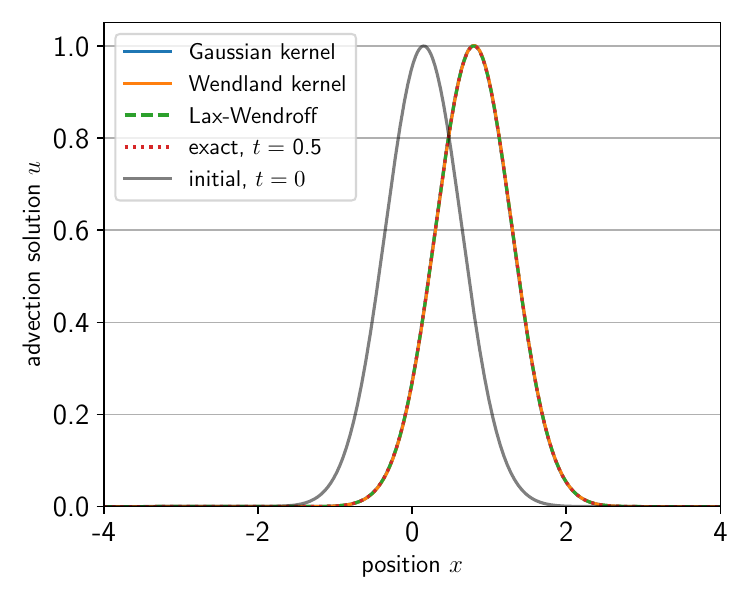}
    \caption{}
    \label{fig:advection2}
    \end{subfigure}
\caption{Solutions of the advection equation with quantum SPH method, with comparison against the classical Lax-Wendroff method and analytical solution. System evolves from initial time $t=0$ to $t=0.5$.}
\end{figure*}

We solve the advection equation in its SPH form (eq. \ref{eq:advection_sph0}). For simplicity, we take a Gaussian profile as initial state, 
\begin{equation}\label{eq:gaussian_ic}
u(x,0) = e^{-(x-\beta)^2/L^2}.
\end{equation}
and define parameters $\beta$ and $L$. Ideally there should be no dispersion so the shape of the Gaussian profile should be perfectly preserved. Hence the analytical solution is simply eq. (\ref{eq:gaussian_ic}) shifted along the x-axis according to time $t$ at speed $c$,
\begin{equation}
u(x,t) = e^{-(x-\beta-ct)^2/L^2}.
\end{equation}

The quantum SPH method considers both Gaussian and Wendland kernels with parameter values outlined in table \ref{tab:advection-table} with constant $\nu=\max(|\nabla W|)$. The simulation is set up so that the domain boundary interactions do not need to be considered. The boundaries are far enough that we do not need to specify boundary conditions in the numerics. We let the system evolve from time $t=0$ to $t_f=0.5$ and we obey the CFL condition (eq. \ref{eq:cfl-advection}) when setting the intermediate timesteps. We compare these solutions with the analytical solution and classical Lax-Wendroff method \cite{Lax60} by calculating the RMS error (eq. \ref{eq:rms}).

The smoothing length $h$ controls the smoothing interpolation error and determines how many SPH particles influence the interpolation for a particular point. Table \ref{tab:advection-table} shows how decreasing $h$ (while fixing the number of SPH particles) increases the solution accuracy. In classical algorithms, the CFL condition determines whether the numerics remain stable and subsequently converge to a solution: the algorithm is successful when $c\frac{\Delta t}{\Delta x} \leq 1$. In the Lax-Wendroff method, decreasing $c\frac{\Delta t}{\Delta x}=1$ to $c\frac{\Delta t}{\Delta x}=0.1$ slightly improves the solution accuracy. In the quantum SPH algorithm, we found that $c\frac{\Delta t}{\Delta x}=1$ can give far less accurate results than when using $c\frac{\Delta t}{\Delta x}=0.1$. This means the quantum algorithm requires noticeably  smaller $\Delta t$ to converge to the most accurate solution, at the expense of iterating over more timesteps and hence longer runtimes. The quantum SPH simulation required a few seconds versus roughly 40 minutes to complete in Fig. \ref{fig:advection1} and Fig. \ref{fig:advection2} respectively. For both tests, the classical Lax-Wendroff method required under two seconds.

\subsection{Solving the diffusion equation}

\begin{figure*}[ht!]
\centering
    \begin{subfigure}{0.49\textwidth}
    \centering
    \includegraphics[width=\linewidth]{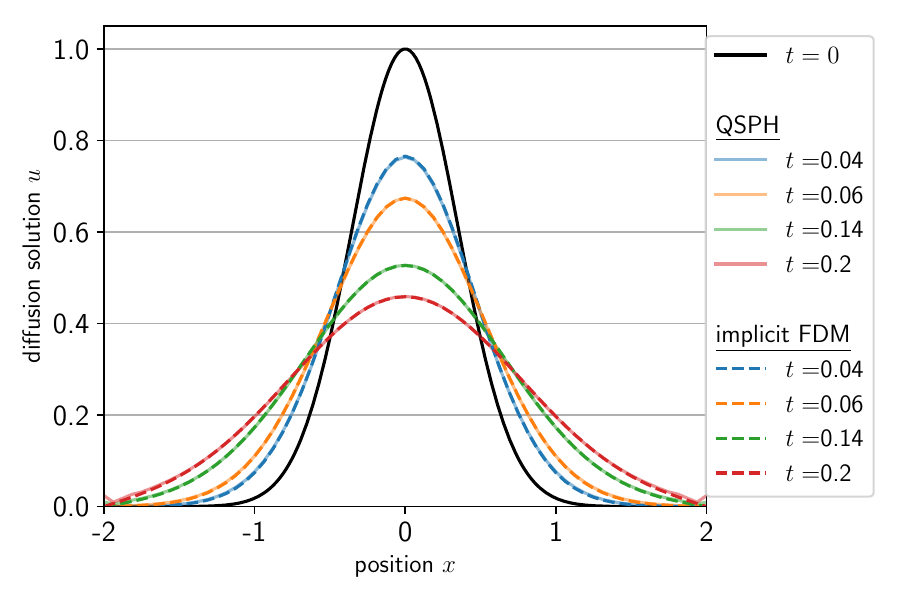}
    \caption{QSPH is quantum SPH method (solid lines), implicit FDM is classical (dashed lines).}
    \label{fig:diffusion}
    \end{subfigure}
\hfill
    \begin{subfigure}{0.47\textwidth}
    \centering
    \includegraphics[width=\linewidth]{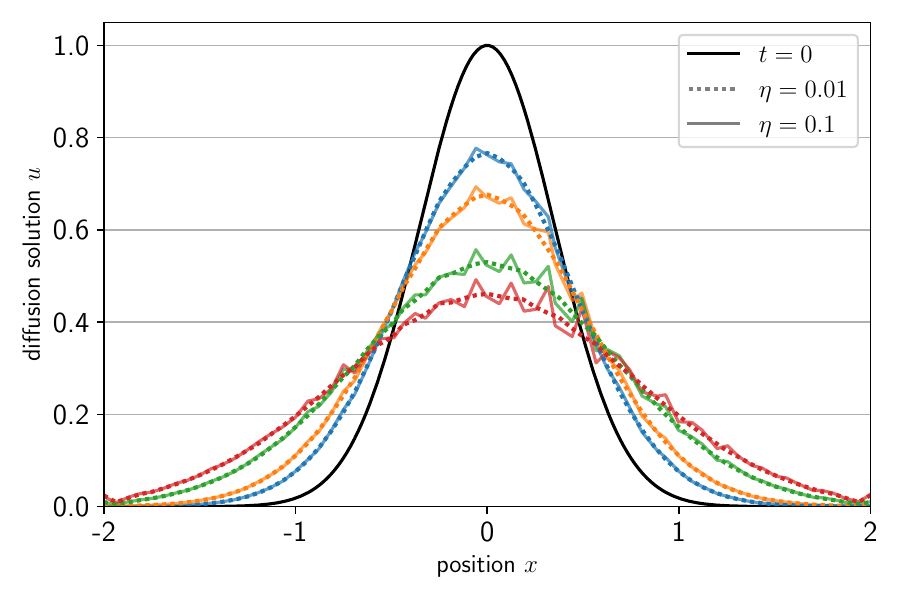}
    \caption{Using QSPH, we vary the SPH particle positions where $\eta$ indicates their displacement from uniform positions in Fig. \ref{fig:domain_discretization}.}
    \label{fig:diffusion-uneven}
    \end{subfigure}
\caption{Solutions of the diffusion equation with Gaussian SPH kernel. Initial state uses Gaussian profile (eq. \ref{eq:gaussian_ic}) with $[\beta,L]=[0,0.5]$. We use $m=6$ qubits and diffusivity constant $\kappa=1.2$. Colors in both graphs correspond to solution at times $t=\{ 0.04, 0.06, 0.14, 0.2 \}$.}
\end{figure*}

The diffusion equation is defined as
\begin{equation}\label{eq:diffusion}
\frac{\partial u(x,t)}{\partial t} - \kappa \frac{\partial^2 u(x,t)}{\partial x^2} = 0
\end{equation}
with diffusivity constant $\kappa$. Following similar arguments above, the SPH form of this PDE is 
\begin{equation}\label{eq:diffusion_sph}
{u}_i^{(n+1)} = 
{u}_i^{(n)} + \kappa\Delta t \sum_{j=1}^N 
(u_j^{(n)}-u_i^{(n)}) \Delta x_j \nabla_i^2 W_{ij}(h).
\end{equation}
The discretized quantum form is 
\begin{equation}\label{eq:diffusion_sph0}
u_i^{(n+1)} = u_i^{(n)} + \kappa\Delta t \; \nu N \norm{\mathbf{a}} \Re \langle a \vert \nabla_i^2 W \rangle.
\end{equation}
The diffusion time step constraint is 
\begin{equation}\label{eq:cfl-diffusion}
\Delta t \leq \frac{(\Delta x)^2}{2\kappa}.
\end{equation}

Figure \ref{fig:diffusion} shows the results of solving eq. (\ref{eq:diffusion_sph0}). To compare, we solve the diffusion PDE (eq. \ref{eq:diffusion}) using the classical implicit finite-difference method (FDM) \cite{Liu09}. There is excellent agreement between the methods. 

As time $t$ increases (Fig. \ref{fig:diffusion}), the initial wave form (black) becomes shorter and wider. This is expected behavior. However there are hints of issues at the boundaries: the `QSPH' solution at $t=0.2$ starts increasing in the limits $x \to -2^+$ and $x \to 2^-$. Unlike in the advection example above, the wave form approaches the boundaries which causes the unusual behavior. This is because we have effectively used constant zero-valued dummy boundary particles at our domain edges by not explicitly encoding a boundary condition in the formulation. More advanced boundary conditions (e.g. mirror or outflow) would resolve such issues. Hence developing a more advanced boundary condition within this quantum framework is the subject of further work.

As an aside, the implicit FDM uses backward Euler time scheme to solve a large, sparse tridiagonal matrix. Recent work \cite{Childs21} has shown that it is possible to use a quantum method based on the HHL algorithm to solve such a matrix. The authors demonstrate its usage on Poisson's equation in two dimensions.

For the QSPH scheme, we also displace the SPH particles from their original positions shown in Fig. \ref{fig:domain_discretization}. Even though we still fix the particles on the 1D line, they are no longer equally spaced. We define these shifted positions as
\begin{equation}
x_k \to x_k + \eta \frac{x_N - x_0}{N} \mathcal{N}(\mu=0,\sigma^2=1)
\end{equation}
with scaling factor $\eta$ and sample taken from the standard normal distribution $\mathcal{N}(\mu=0,\sigma^2=1)$. When the SPH particle displacement is small ($\eta=0.01$), the solution is smooth and well-behaved (Fig. \ref{fig:diffusion-uneven}) but does not have the excellent agreement seen in Fig. \ref{fig:diffusion}. As we increase the scaling factor to $\eta=0.1$, the solution becomes much noisier especially as time increases. This is expected as the SPH accuracy decreases significantly for irregular distributions. However we demonstrate that the algorithm still works by using unequally spaced particles. This is closer to the general SPH problem formulation -- it is a step towards a more complicated setup where we keep track of freely-moving SPH particles at each time step.

We note that to calculate the second derivative of $u$, we directly take the second gradient of the kernel as opposed to using the Laplacian operator discussed in Morris et al.'s work \cite{MORRIS1997214}. This choice may also explain the noise observed in Fig. \ref{fig:diffusion-uneven} for increased particle disorder, as the direct second kernel gradient is known to introduce oscillatory behavior. Implementing a quantum discretization of the Morris Laplacian is indeed possible using this framework. However the test cases in this work offer a fundamental and in-principle demonstration of our quantum discretization approach. Investigations into the many different SPH operators (first gradients and Laplacians) under this quantum framework will be the subject of future work. This analysis would involve more practical two-dimensional fluid flow test cases where the use of different SPH operators has greater benefit.

\section{Discussion}
\label{sec:Discussion}

The numerical simulations illustrate the potential power of using quantum computers to perform SPH calculations. When calculating a function's derivatives, a small increase in the number of qubits in the quantum register allows the computation to contain many more SPH particles and therefore significantly increase the accuracy. This is seen in Fig. \ref{fig:function} and \ref{fig:deriv1} when comparing the $m=4$ and $m=8$ approximations. For the second derivative approximation (Fig. \ref{fig:deriv2}), the increase in accuracy with the quantum register size is still evident despite the approximation being somewhat less accurate than for the first derivative and the function itself. The RMS error for approximating eq. \eqref{eqn:Runge_function} decreases rapidly as we increase the register size (Fig. \ref{fig:rms_error}). The function approximation (and derivatives) continues to increase in accuracy as the register size increases until we reach the SPH discretization error limit.

\begin{figure}[ht!]
\centering
\includegraphics[width=\linewidth]{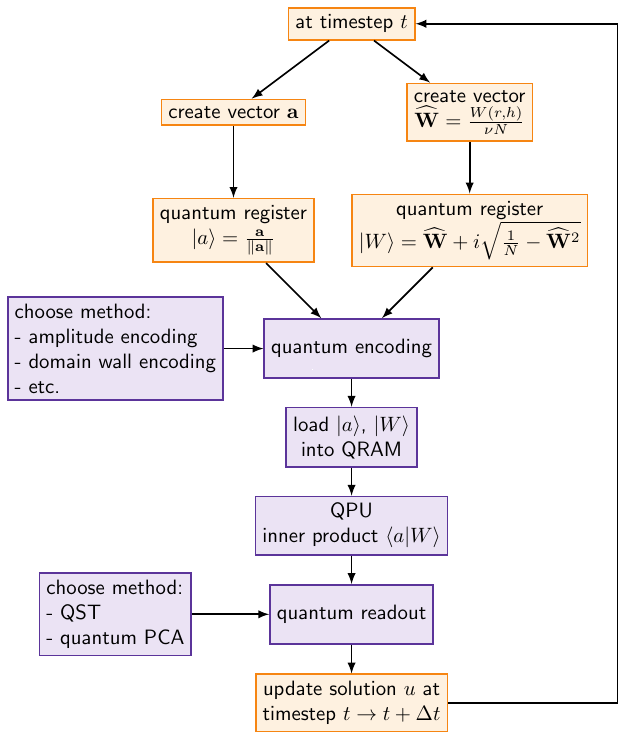}
\caption{Schematics of quantum algorithm. Classical (quantum) procedures in orange (purple). Note that we simulate the `quantum' steps on a classical computer.}
\label{fig:flowchart}
\end{figure}

Similarly, our method can solve the advection and diffusion equations, two fundamentally important PDEs that underpin many physical processes across science and engineering applications. We show the schematics of our proposed algorithm in Fig. \ref{fig:flowchart}, outlining the classical and quantum subroutines. The classical parts involve preparing the quantum registers and updating the solution $u$ at each timestep. Both are computationally expensive. The (anticipated) quantum computer contains a quantum RAM (QRAM) and quantum processing unit (QPU) to calculate the inner product. This is book-ended by a quantum encoding and readout, both potentially expensive procedures that transfer quantum information between the classical and quantum hardware. Hence we have shown a successful proof-of-concept for a quantum SPH method. The next step is increasing its efficiency.

The method presented in section \ref{sec:derivatives} is general. It allows for any kernel function (including derivative functions) to be used in the approximation and for arbitrary domains, functions and register sizes. By encoding the function, spatial discretization and kernel function into a quantum register, it is possible to significantly increase the number of SPH particles. Although our results imply that any quantum advantage relies on some efficient method for creating the encoded registers $\ket{a}$ and $\ket{W}$, the reality is that such a method may not exist. If there are exponentially many classical values of $a_j$ and $W_j$ in the state, then it is faster to calculate the inner product using multiplication and addition. An efficient method for generating the quantum registers would only be useful when those states are prepared from a previous quantum process that does not use exponentially many amplitude values to prepare the state. This is relevant for example, if we devise a method to iterate over the timesteps using only quantum methods.

\subsection{SPH particle positions}

When constructing the quantum register in section \ref{sec:derivatives}, we fix the SPH particles in space for simplicity - they remain in position $r_k$, but can have non-uniform or uniform size $\Delta x_k$ (Fig. \ref{fig:domain_discretization}). In comparison, classical SPH is a Lagrangian method where the particles can move freely in space. Future work will involve generalizing our method to account for different particle locations to be more aligned with the classical SPH formulation. This would potentially introduce another quantum register to store the extra degree of freedom. Classical SPH uses special neighbor-searching subroutines to efficiently calculate the kernel $W_{i,j}$ and solutions. Hence we must also use a neighbor-search quantum equivalent to minimize any numerical bottlenecks in Fig. \ref{fig:flowchart}. There are numerous available algorithms that may be adapted for SPH neighbor searching. For example Grover's landmark search algorithm \cite{grover1996fast} provides a database search in $O(\sqrt{N})$ (over $N$ entries). Grover search can also be implemented as a quantum walk algorithm \cite{Kendon06,Ambainis03}. Both could offer an effective search method when combined with existing SPH neighbor-list approaches, eg. cell-linked or Verlet lists \cite{dominguez2011neighbour}.

\subsection{Inner product}

In Fig. \ref{fig:flowchart}, we calculate an inner product of two quantum registers using brute force vector multiplication. One alternative is to use the swap test \cite{Barenco97,Buhrman01} or one of its variations \cite{Fanizza20} to speed up the calculation. The swap test combines quantum phase estimation algorithm and Grover searching to find the probability of some desired quantum state. Compared to classical algorithms, the swap test achieves exponential acceleration. However it must be repeated multiple times and each measurement result is used to approximate the probability which partly offsets the quantum speedup. Alternatively, we may use a quantum algorithm for approximate counting with variants available with \cite{Brassard98} and without \cite{Aaronson19} using a quantum Fourier transform.

For a more efficient implementation on NISQ devices, we may use the Bell-basis algorithm \cite{Cincio18}, a constant circuit-depth algorithm for computing state overlap. It has significantly lower error and better scaling than the swap test (linear scaling). From the machine learning toolbox, quantum mean estimation and support vector machine are used to calculate the state overlap \cite{Liu22}.

\subsection{Time stepping}

It is computationally expensive to use classical time stepping when solving the advection (eq. \ref{eq:advection_sph0}) and diffusion equations (eq. \ref{eq:diffusion_sph0}). In these recurrence relations, the solution $u$ at timestep $n$ relies on the solution at previous timesteps (Fig. \ref{fig:flowchart}). We use classical for-loops to iteratively find solution $u$ at a final time $t_f$. This is a major bottleneck because Fig. \ref{fig:flowchart} implies that the quantum registers must be rebuilt for each timestep. In addition, the difficulty in converting the classical data into quantum registers (and vice versa) is the same order of magnitude as the timestepping.

Ideally we want to perform the calculation for multiple time points simultaneously using a single quantum operator. Alternatively, we could develop a method for classical data-to-quantum conversion at the start of the algorithm, and quantum-to-classical to output the result at the final time. Including a quantum timestepping subroutine could significantly reorganize the procedure shown in Fig. \ref{fig:flowchart}. However the core algorithmic components such as taking the inner product and quantum encoding procedure should remain the same.

\subsection{Quantum encoding and readout}

Programming quantum computers is challenging due to their quantum nature and hardware limitations. One key difference to classical computing is how we handle the data. Quantum computers do not currently have access to databases or quantum version of RAM \cite{Giovannetti08}. Therefore we load data into quantum computers by encoding it into the qubit state. There is no broad consensus on how best to encode classical data into qubits before loading into a quantum random access memory (QRAM).

The research community considers quantum encoding one of the grand challenges of building viable quantum computers. In any quantum computation, a fast algorithm for initializing the quantum data is critically important for reducing the runtime. The encoding procedure should be designed with quantum circuit elements since the circuit model provides systematic and efficient instructions to achieve universal quantum computation. Current devices contain error-prone quantum gates and a limited number of qubits with short decoherence times. Hence encoding methods are a trade-off between two major factors: number of required qubits and runtime complexity. In addition to the quantum no-cloning theorem \cite{Wootters82}, these factors dominate the overall computational cost due to the quantum measurement postulate: we often repeat the same algorithm many times to retrieve measurement statistics while each measurement destroys the quantum state. In the worst case, loading the data requires exponential time. To efficiently encode a large amount of data, a logarithmic or linear runtime is still ideal. 

There are numerous ways to represent classical data in a Hilbert space (Fig. \ref{fig:flowchart}). In section \ref{sec:Quantum_registers}, we introduced the idea of quantum registers which is amenable to quantum amplitude encoding. This method represents a data vector by the amplitudes of a quantum state \cite{Long01}. The embedding uses fewer qubits than other methods like basis encoding and Hamiltonian encoding, making it ideal for NISQ era devices where qubits are in short supply \cite{Bharti22,Preskill18}. One example of amplitude encoding uses quantum Fourier transforms (QFTs) \cite{Mottonen05}. It is possible to use the method in the Grover-Rudolph \cite{Grover02} approach to load probability distributions and hence efficiently encode polynomials \cite{Gilliam21} and binary strings (zero amplitudes are zeroes and non-zero amplitudes are ones). The divide-and-conquer algorithm \cite{Araujo21} uses controlled swap gates and ancilla qubits. It recursively breaks down the encoding problem into many sub-problems so that they are simple enough to be solved directly, before recombining to form the whole register. Another option is to use a quantum random access memory (QRAM) architecture comprising flip-flop QRAM (FF-QRAM) procedures to register the classical data structure into quantum format \cite{Park19,deVeras21}. Domain-wall encoding is a highly active area of research in quantum annealing and optimization \cite{Chancellor19,Berwald23} which can also be applied to our problem.

After performing the swap test on QPU, we want to output values to the classical computer and update the solution $u$ at time $t+\Delta t$ (Fig. \ref{fig:flowchart}). The aim of quantum state tomography (QST) is to estimate an unknown quantum state when many identical copies are available so that we can perform different measurements on each copy \cite{DAriano03,Christandl12}. Homodyne tomography is an early example that reconstructs the density matrix $\rho$ of an unknown state, however it is too computationally expensive for practical problems. Further improvements include the ``maximum likelihood, minimum effort'' method which introduces an optimization algorithm to increase the fidelity \cite{Smolin12}. Alternatively, a machine learning approach uses a variational algorithm and swap test as cost function \cite{Liu20}. Another option is quantum principal component analysis (PCA) \cite{Lloyd14}. This uses density matrix exponentiation and quantum phase estimation to provide the eigenvectors of $\rho$. It is simpler and faster than other strategies for performing entangled measurements on many copies of $\rho$ such as the quantum Schur transform \cite{Bacon06}.

\subsection{Benchmarking procedure}

To show that the quantum register scheme is useful, we need a robust benchmarking procedure to quantify any quantum advantage. It is possible that our method provides an advantage for three-dimensional systems, whereas one-dimensional systems are more useful for developing the method. This is an important issue to consider in the long term.

\section{Conclusions and future work}
\label{sec:Conclusion}

This work has shown a scheme for encoding the SPH approximation method in a quantum register. We demonstrated classical simulations of the quantum scheme for both the Gaussian and Wendland kernel functions using different registers sizes to approximate a function and its first and second derivatives. This scheme demonstrates that the error in the approximation decreases exponentially with the number of qubits in the register. 

Quantum computing promises to revolutionize many scientific fields and none more so than numerical analysis and computational modelling. A method combining SPH and quantum computation could allow us to perform accurate continuum mechanics simulations with complex geometries for problems which are currently intractable.

\subsection{Future work}

There is much to do before the method presented in our work can have any practical uses. There are two broad areas of research opportunities that we classify as short- and long-term projects. 

In the short term, we can extend the method to two- and three-dimensional systems, then demonstrate its use in science and engineering problems. Other research avenues involve incorporating more general boundary conditions in the algorithm; checking that the method works in the presence of significant non-linearity; exploring other initial solutions that are not as well-behaved as Gaussian profiles; allowing the (currently fixed) SPH particles to move freely in space; performing stability analysis; and developing benchmarking techniques. Of course in the short term, some of these should show the same behavior as classical SPH, while we still work classically. How they look with the inner product done on a QPU is of course different and interesting, and essential to investigate in the longer term.

In the long term, we must address the issue of quantum encoding, as discussed above. Our method uses classical means to construct the quantum registers $\ket{a}$ and $\ket{W}$, and simulate the SPH approximation at consecutive timesteps (Fig. \ref{fig:flowchart}). Finding quantum alternatives for these subroutines is ideal, as well as using a quantum method to calculate the inner product $\langle a \vert W \rangle$ and to convert the quantum outputs into classical information. It is possible that quantum readout is not necessary, especially if the time-stepping can be done in a quantum way. This could allow our method to be fully quantum, so that only the initial time $t=0$ encoding and final time $t_f$ decoding would require converting between classical and quantum domains (Fig. \ref{fig:flowchart}).

\section*{Author contributions}

\textbf{R. A. \orcidlink{0000-0002-0082-5382}:} Methodology, Software, Formal analysis, Investigation, Writing - Original Draft, Writing - Review \& Editing, Visualization.

\textbf{A. J. W.:} Methodology, Software, Formal analysis, Investigation, Writing - Original Draft.

\textbf{V. M. K. \orcidlink{0000-0002-6551-3056}:} Methodology, Resources, Supervision, Project administration, Funding acquisition.

\textbf{S. J. L. \orcidlink{0000-0001-9701-6524}:} Methodology, Conceptualization, Resources, Supervision, Project administration, Funding acquisition

\section*{Funding}

R. A. and V. M. K. are supported by UK Research and Innovation (UKRI) Grants EP/T001062/1, EP/W00772X/2, EP/T026715/2, and EP/Y004566/1. S. J. L. is supported by UKRI Grants EP/R04189X/1 and EP/Y004663/1.

\section*{Declaration of Interests}

The authors declare that they have no known competing financial interests or personal relationships that could have appeared to influence the work reported in this paper.

\bibliographystyle{elsarticle-num}
\bibliography{main}

\end{document}